\documentclass[showpacs,letterpaper,aps,pra,twocolumn,floatfix,superscriptaddress]{revtex4}
\usepackage{graphicx}
\usepackage{amssymb}
\usepackage{epstopdf}
\usepackage{amsmath,amsfonts,latexsym}
\newcommand*{\brp}{\mathbf{r}'}
\newcommand*{\br}{\mathbf{r}}
\newcommand*{\bp}{\mathbf{p}}
\newcommand*{\bk}{\mathbf{k}}
\newcommand*{\bq}{\mathbf{q}}

\newcommand*{\cC}{{\cal C}}

\newcommand*{\cH}{{\cal H}}

\newcommand{\psid}{\psi^{\dagger}}
\newcommand{\psipd}{\psi^{\vphantom{\dagger}}}

\newcommand{\be}{\begin{equation}}
\newcommand{\ee}{\end{equation}}
\newcommand{\bea}{\begin{eqnarray}}
\newcommand{\eea}{\end{eqnarray}}
\newcommand{\ba}{\begin{align}}
\newcommand{\ea}{\end{align}}
\newcommand{\TIm}{\textrm{Im}}
\newcommand{\TRe}{\textrm{Re}}
\newcommand{\omeps}{\bar{\omega}}

\newcommand{\uu}{1}
\newcommand{\dd}{2}
\newcommand{\eF}{{\epsilon}_{\rm F}}
\newcommand{\kFS}{k_{{\rm F}\sigma}}
\newcommand{\kF}{{k_{\rm F}}}
\newcommand{\TU}{V}
\newcommand{\s}{\sigma}
\newcommand{\si}{{\sigma^\prime}}
\newcommand{\ham}{\cH}

\def\pra{\ref@jnl{Phys.~Rev.~A}}
\def\prb{\ref@jnl{Phys.~Rev.~B}}
\def\prc{\ref@jnl{Phys.~Rev.~C}}
\def\prd{\ref@jnl{Phys.~Rev.~D}}
\def\pre{\ref@jnl{Phys.~Rev.~E}}
\def\prl{\ref@jnl{Phys.~Rev.~Lett.}}

\def\ket#1{|#1\rangle}

\begin{document}

\title{Radio frequency spectroscopy of a strongly imbalanced Feshbach-resonant Fermi gas}
\author{Martin Veillette}
\affiliation{Department of Physics,
Berea College, Berea, KY 40404}
\author{Eun Gook Moon}
\affiliation{Department of Physics, Harvard University, Cambridge, MA 02138}
\author{Austen Lamacraft}
\affiliation{Department of Physics, University of Virginia,
Charlottesville, VA 22904-4714}
\author{Leo Radzihovsky}
\affiliation{Department of Physics, University of Colorado, Boulder, Colorado 80309}
\author{Subir Sachdev}
\affiliation{Department of Physics, Harvard University, Cambridge, MA 02138}
\author{D.~E.~Sheehy}
\affiliation{Department of Physics, Louisiana State University, Baton Rouge, LA 70803-4001}

\date{\today}

\begin{abstract}
  A sufficiently large species imbalance (polarization) in a
  two-component Feshbach resonant Fermi gas is known to drive the
  system into its normal state. We show that the resulting
  strongly-interacting state is a conventional Fermi liquid, that is,
  however, strongly renormalized by pairing fluctuations.  Using a
  controlled $1/N$ expansion, we calculate the properties of this state
  with a particular emphasis on the atomic spectral function, the
  momentum distribution functions displaying the Migdal discontinuity,
  and the radio frequency (RF) spectrum. We discuss the latter in the
  light of the recent experiments of Schunck \emph{et al.} (Science
  {\bf 316}, 867 (2007)) on such a
  resonant Fermi gas, and show that the observations are consistent
  with a conventional, but strongly renormalized
  Fermi-liquid picture.
 \end{abstract}


\maketitle

\section{Introduction}

One of the key recent developments in studies of degenerate atomic
gases is the tunability of atomic interactions via a Feshbach
resonance (FR). This has led to the realization of a resonantly-paired
s-wave superfluid that can be tuned between the two well-studied
limits of a weakly-paired Bardeen-Cooper-Schrieffer (BCS) superfluid and a
strongly-paired diatomic molecular Bose-Einstein condensate (BEC)
superfluid~\cite{regal2004,zwierlein2004,chin2004,bourdel2004,kinast2004,zwierlein2005}.

The two asymptotic superfluid regimes at large positive and negative
FR detunings allow a detailed quantitative description, made possible
by the existence of a small gas parameter, $n a^3$, corresponding to
the ratio of a short scattering length $a$ to a large average particle
spacing $\ell = n^{-1/3}$. In contrast, although the intermediate low
temperature crossover regime is a conventional superfluid, that
smoothly interpolates between the BCS and BEC
limits,
its quantitative description (in a broad
resonance case) is hindered by strong interactions, characterized by a
diverging scattering length and absence of a natural small
parameter \cite{narrowFBR}. The flip side, of course, is that a
diverging scattering length leaves particle spacing as the only
relevant length scale, leading to a universal phenomenology of a
resonant Fermi gas near a unitary point.

While earlier studies focused on the case where the populations of the
two atomic species involved in pairing are equal (vanishing
polarization), and thus on the nature of the superfluid phase, recent
experimental and theoretical investigations have extensively explored
the {\em imbalanced} resonant Fermi gas, extending its FR detuning
phase diagram to a finite
polarization~\cite{zwierlein2006,partridge2006,
zwierlein2006b,shin2006,partridge2006b,bedaque2003,sheehy2006,sheehyAOP2007}.
These studies have consistently found that at high polarization the
state is non-superfluid, and have treated it as a simple
Fermi gas. However, at low detuning
the system is strongly resonantly interacting, and
a detailed description of the highly polarized normal state near the resonance
remains a challenge~\cite{singleFermion}.

Some light on the complex nature of the strongly-interacting normal
state has been shed by a recent radio frequency (RF) spectroscopy
experiment~\cite{schunck2007}.  At high polarization an absence of a
BEC peak on one hand and presence of a temperature-dependent spectral
shift relative to the atomic line, on the other, was
observed. Interpreting the latter as a pairing gap~\cite{chin2004},
taken together, these observations have been interpreted as evidence
for a paired non-superfluid state.  While unsurprising at finite
temperature, and certainly present on the BEC side of the resonance,
where the gap is set by molecular binding energy, an existence of such
a state at {\em zero} temperature would constitute a dramatic
departure from a modern understanding of possible condensed matter
ground states. In particular, the only established route to a
suppression of boson (Cooper pairs or diatomic molecules in the
present context) superfluidity at zero temperature is through a
localization of bosons by quenched disorder into a Bose
glass~\cite{fisherBG}, by a commensurate (e.g., imposed optical)
lattice into a Mott insulator~\cite{fisherBG,greiner2002}, or by
crystallization as in a case of a solid $^4$He. Since none of these
mechanisms appears to be at play in the trapped dilute atomic gas
studied in experiments by Schunck \emph{et al.}~\cite{schunck2007},
their observations and conclusions remain puzzling.

The interpretation of RF spectra in the strongly-interacting regime
has been the subject of numerous theoretical investigations
\cite{kinnunen2004,ohashi2005,yu2006,baym2007,Basu}, but remains incomplete. The
recent RF spectroscopy experiments on a strongly polarized Fermi
gas\cite{schunck2007,shin2007}, have rekindled theoretical studies of
such a system with a focus on the unitarity regime~\cite{lobo2006,bulgac2007,CarlsonReddy} and its strong
interactions in the {\em normal} state~\cite{combescot2007,punk2007,perali2007,he2007,massignan2008,prokofev2008}.

\begin{figure}
\centering \includegraphics[width=0.45\textwidth]{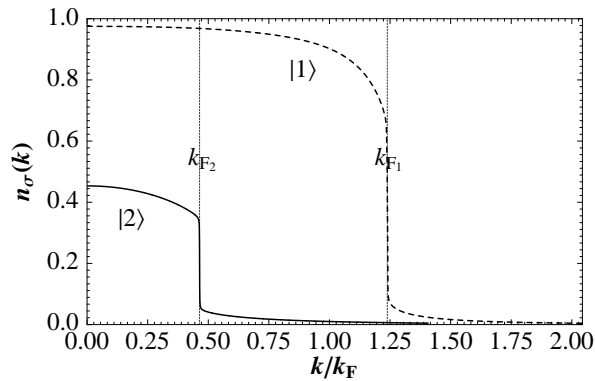}
\caption{Momentum distribution $n_\s(k)$ of the majority ($\ket1$) and minority
  ($\ket2$) atoms for polarization $P=0.9$ at zero temperature and
  at resonance. The residues for the majority and minority atoms
  are, respectively, $Z_1=0.56$ and $Z_2=0.29$.}
\label{fig:nk}
\end{figure}

The purpose of the present paper is to make a case for a more conservative 
interpretation of the RF spectra of Refs.~\onlinecite{schunck2007,shin2007}, 
namely that a non-superfluid state of a
highly polarized, resonantly interacting Fermi gas is in fact a Fermi
liquid, albeit a strongly renormalized one. Indeed, the Luttinger
relations discussed in Refs.~\onlinecite{powell,kunyang}
require any non-superfluid state to have
Fermi surfaces for both the majority and minority species, enclosing
the same volumes as for non-interacting fermions.
We will show that such a Fermi liquid is
perfectly consistent with the experiment of
Ref.~\onlinecite{schunck2007}, and that the observed shift in the RF
spectrum of the minority atoms is the result of large self-energy
effects due to the Feshbach resonant scattering, as also emphasized in
parallel recent studies~\cite{punk2007,perali2007,he2007}. Based on
this we suggest that a far better test of the nature of the ground
state is a measurement of the momentum distribution $n_{\dd}(k)$ of
the minority atoms; in the following the label $\s=\uu,\dd$ denotes
the majority and minority species, respectively. Such a measurement
would be a direct test for the existence of a Fermi surface, marked by
a Migdal discontinuity and characterized by the quasiparticle residues
$0<Z_\s<1$
\be
Z_\s \equiv n_\s(\kFS-)-n_\s(\kFS +),
\ee
illustrated in Fig.~\ref{fig:nk}. This hallmark Fermi liquid feature
would be absent in the case of a paired ground state, allowing for a
sharp qualitative distinction between two possibilities.
An interesting feature of Fig.~\ref{fig:nk} is that the functions
$n_\sigma (k)$ are both universal as functions of $k/k_F$
at resonance {\em i.e.\/} for the
Fermi gas at unitarity, the functional forms (including the values of
the discontinuities $Z_\sigma$) depend {\em only\/} upon the polarization
$P$. Our computations of these universal functions is however not exact,
and next we describe our computational method.

We build on our earlier studies of the superfluid
state\cite{nikolic2007,veillette2007}: our theoretical approach to the
treatment of this strongly interacting (small parameter-free) normal
state is based on the introduction of an artificial small parameter
$1/N$, with $N$ the number of distinct ``spin''-1/2 fermion flavors in
the generalized model.  The advantage of such a generalization is
that, for $N \rightarrow \infty$, the problem is exactly solvable,
with finite $N$ corrections computable via a systematic expansion in
$1/N$ about this solvable limit.  In particular, we calculate the RF
spectrum to  subleading order in $1/N$ using the full interaction
matrix, including the appropriate vertex correction. We then use these
controlled $1/N$ results to extrapolate to the experimentally-relevant
case of a single flavor ($N=1$) of two opposite-``spin'' (hyperfine
levels) fermionic atoms.
This extrapolation is a subtle issue, and we cannot rule out the
possibility of a nonanalyticity in the $N\to 1$ limit.  However,
such nonanalyticities are rare and there is no reason to expect
them here.  

\begin{figure}
\centering \includegraphics[width=0.45\textwidth]{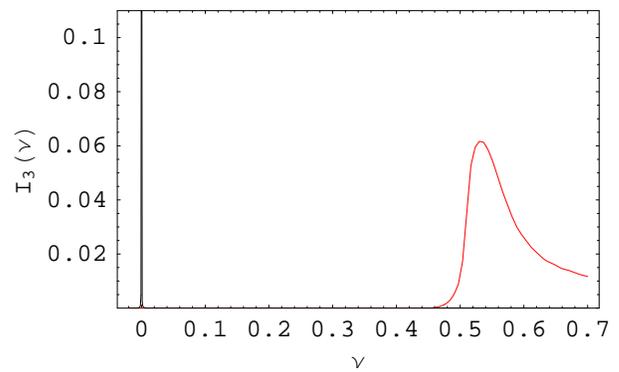}
\caption{RF spectrum for polarization $P=0.97$:
 the intensity
$I_3(\nu)$ (arbitrary units)
  vs. the detuning from the resonance frequency $\nu$ measured
  in units of the Fermi energy $\eF$. In free space, the resonance would be at
  $\nu=0$. The shift in the resonance frequency above is due to strong interactions
  between fermions in the non-superfluid ground state
  of a polarized Fermi gas. Our primary claim is that such a shift is present
  even while the ground state remains a Fermi liquid, with the discontinuities
  in the momentum distribution function shown in Fig.~\ref{fig:nk}.}
\label{fig5}
\end{figure}

We regard the large-$N$ expansion as  
providing a framework for understanding qualitative aspects of 
the renormalized Fermi liquid properties of 
strongly-interacting imbalanced Fermi gases.  
A sample result for the RF spectrum obtained via this approach 
is shown
in Fig.~\ref{fig5}.
Note the asymmetric lineshape: this arises from the imaginary part of 
the pairing fluctuations propagator which contains the phase space for
the decay of a Cooper pair into a two fermion final state.

The universality considerations noted above for $n_\sigma (k)$ also apply to
the the RF absorption spectrum. However, here there are also additional complications
associated with the scattering lengths of excited states, and these will
be discussed below.

The outline of the paper is as follows.  In Sec. \ref{formalism} we
introduce the resonant single-channel model as well as its $N$-flavor
generalization, and carry out its systematic expansion in $1/N$. Then,
in Sec. \ref{fermi}, we illustrate the Fermi liquid properties of the
system at high polarization. In
Sec. \ref{radio} the radio-frequency probe is described and we discuss
our results in terms of the recent experiments of Schunck \emph{et al.},
and conclude with a brief summary in Sec. \ref{conclusion}.

\section{Formalism}
\label{formalism}
\subsection{Model}
\label{model}
To capture the physics involved in the RF experiment, we model the
system in terms of the three lowest hyperfine states,
$\ket\sigma=\ket1,\ket2$ and $\ket3$. The two lowest states, $\ket1$
and $\ket2$, are loaded with atoms and responsible for the strong
superfluid correlation, while the higher hyperfine state, $\ket3$, is
initially empty. Experimentally, the RF field at frequency
$\omega$ is used to induce atomic transitions from the state $\ket2$
to the state $\ket3$, and the induced transition rate is measured as a
function of $\omega$. As we will show below, this allows one to
experimentally probe a two-particle correlation function.

Although our consideration for RF spectroscopy on fermionic atoms will
be quite general, we will put particular emphasis on recent
experiments on $^6$Li. In this system, the three lowest lying states,
$\ket1, \ket2$ and $\ket3$ can be identified with
$|F=1/2,m_F=1/2\rangle$, $|F=1/2,m_F=-1/2 \rangle $ and
$|F=3/2,m_F=-3/2\rangle$, respectively.

By integrating out the higher order hyperfine states we obtain an
effective Hamiltonian in terms of the three lowest states, given by
\begin{align}
\ham= &\sum^3_{\s=1}
\int d^3 \br \; \psid_\s (\br) \left( -\frac{\nabla^2_{\br}}{2 m} +
  \omeps_{\s} -\mu_\sigma
\right) \psipd_\s (\br) \notag \\
& +  \frac{1}{2} \sum^3_{\s,\si=1} \int d^3 \br  \; \lambda_{\s \si}
\psid_\s(\br) \psid_\si(\br)
 \psipd_\si(\br) \psipd_\s (\br), \label{m1.1}
\end{align}
where $\lambda_{\s \si}$ are couplings (interaction strengths set by
the corresponding scattering lengths $a_{\s \si}$) between states
$\ket\s$ and $\ket\si$. $\psid_\s(\br), \psipd_\s(\br)$ are,
respectively, the fermion creation and annihilation operators at
position $\br$ and hyperfine state $\s$, which obey the usual
anticommutation relation $\left\{ \psipd_\s (\br), \psid_\si (\brp)
\right\} = \delta(\br-\brp) \delta_{\s,\si}$.  The detuning of the
level $\s$ is controlled by a Zeeman field encoded by the detuning
parameter $\omeps_\s$.  The chemical potential $\mu_\s$ fixes the
average atom density $n_\s$ in the hyperfine state (spin) $\s$.

While in above model the particle number in each hyperfine state is a
good quantum number, in principle there are additional interaction
channels present in the physical system, that break this symmetry.
For instance \cite{40Kstates} in $^{40}K$, (where $\ket1, \ket2$, and $\ket3$ are
$|F=9/2, m_F=-9/2 \rangle , |F=9/2, m_F=-7/2 \rangle$ and $|F=9/2,
m_F=-5/2 \rangle $) a collision term such as $\psid_1 \psid_3 \psi_2
\psi_2$ conserves the total quantum number $m_F$ is therefore allowed
by symmetry. However, effects of such interactions, that do not
conserve the atom number in each hyperfine state are energetically
suppressed by virtue of $2 \omeps_2 \neq \omeps_1+\omeps_3$, due to
the large quadratic Zeeman splitting between the three levels.
Experimentally, the Feshbach resonances are chosen such that these kind
of non-conserving processes are minimized, and therefore we will not
consider them any further here.

For a short-range s-wave interaction, the Pauli principle enforces
$\lambda_{\s \s}=0$, which together with the exchange symmetry
$\lambda_{\s \si}=\lambda_{\si \s}$, reduces the nine coupling
constants $\lambda_{\s \si}$ down to three. The corresponding three
two-particle scattering lengths $a_{\s \si}$ for atoms in states
$\ket\s$ and $\ket\si$ (for $\s \neq \si$) are related to the
strengths of the couplings $\lambda_{\s \si}$ via the relation
\be \frac{m}{4 \pi a_{\s
\si}}= \frac{1}{\lambda_{\s \si}} + \int \frac{d^3 \bk}{(2 \pi)^3}
\frac{1}{2 \epsilon_\bk},
\label{m1.3}
\ee
where $\epsilon_\bk=k^2/(2m)$ is the the free fermion dispersion.

\subsection{Large $N$ expansion}

As discussed in the Introduction, because we are interested in the
system in the vicinity of the unitary point, where the atom
interaction is strong, an analysis based on a straightforward
perturbation theory in the gas parameter $n a^3$ (that diverges at the
unitary point) clearly fails and an approach nonperturbative in $n
a^3$ is required. To this end, we employ the procedure, introduced and
successfully utilized for the superfluid phase in
Refs.~\onlinecite{nikolic2007,veillette2007}, of generalizing the
physical model to that of $N$ flavors of each of the three hyperfine
states species, with the Hamiltonian given by
\begin{align}
\ham&= \sum^3_{\s=1} \sum_{i=1}^N
\int d^3 \br \; \psid_{i \s} (\br) \left( -\frac{\nabla^2_{\br}}{2 m} +
  \omeps_{\s} -\mu_\sigma
\right) \psipd_{i \s} (\br) \notag \\
&+  \frac{1}{2 N} \sum^3_{\s,\si=1} \sum_{i,j=1}^N\int d^3 \br  \;
\lambda_{\s \si} \psid_{i \s}(\br) \psid_{i \si}
(\br) \psipd_{j \si}(\br) \psipd_{j \s} (\br).
\label{m1.1N}
\end{align}

It is then straightforward to develop an expansion in powers of $1/N$,
taking the physically relevant limit $N=1$ at the end of the
calculation.  Since we are interested in the normal properties only,
the full effective action formalism of
Refs.~\onlinecite{nikolic2007,veillette2007} will not be required.
Instead we may calculate the $1/N$ contributions diagrammatically,
observing that vertices bring factors of $1/N$, and particle-particle loops a factor of
$N$, as we sum over all components. Having thereby established the
relevant class of leading $1/N$ diagrams, for ease of notation, we
then drop the $i$ flavor index.

In this way, we arrive at the thermal Green's function of species $\s$
defined by
\be G_\s(\bk,i \omega_n)= - \int_0^\beta d \tau e^{i
\omega_n \tau} \langle T_\tau \left\{ \psipd_{\s}(\bk,\tau) \psid_{\s}
(\bk,0) \right\} \rangle,
\ee
where $\omega_n=2 \pi \left(n+\frac12 \right)/\beta$ is the fermionic
Matsubara frequency and $\beta=1/T$ is the inverse temperature $T$ and
$T_\tau$ is the time ordering operator with the imaginary time
$\tau$. As in the standard perturbation theory, we express the Green's
function in term of its non-interacting form and the self-energy
\be G^{-1}_\s(\bk,i \omega_n)= G^{(0) -1}_\s(\bk, i \omega_n) -
\Sigma_\s(\bk,i \omega_n),
\ee
where $\Sigma_\s(\bk, i \omega_n)$ is the (Matsubara) self-energy,
$G^{(0)}_{\s}(\bk,i \omega_n)=\left(i \omega_n-\varepsilon_{\s,\bk}
+\mu_\s \right)^{-1}$ is the bare Green's function, and
$\varepsilon_{\s,\bk}=\epsilon_\bk +\omeps_\s$ is the bare fermionic
dispersion.

\begin{figure}
\centering \includegraphics[width=0.5\textwidth]{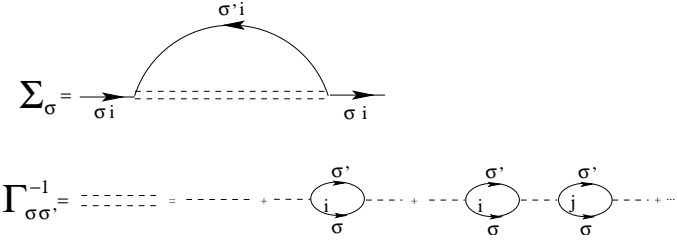}
\caption{Self-energy diagram $\Sigma_{\s \si}$ to leading order in
$1/N$.}
\label{fig:diagram}
\end{figure}

The self-energy is determined to lowest order in $1/N$. It can be
expressed as
\begin{align}
\Sigma_\s(\bk,i\omega_n) \equiv \sum^3_{\si=1(\neq \s)} \Sigma_{\s
  \si}(\bk, i\omega_n)
\end{align}
where $\Sigma_{\s \si}$ is the self energy contribution of the level
$\ket\s$ due to interactions with level $\ket\si$, corresponding to
the diagram of Fig.~\ref{fig:diagram} and given by
\begin{align}
\label{SE}
\Sigma_{\s \si}(\bk,i \omega_n)=
&-\frac{1}{\beta N} \sum_{\Omega_m}\int
\frac{d^3\bq }{\left(2\pi\right)^3}
\Gamma_{\s \si}(\bq,i\Omega_m)\\
&\times G^{(0)}_{\si}(\bq-\bk,i\Omega_m-i\omega_n).
\notag
\end{align}
The large-$N$ renormalized interaction vertex (the T-matrix) is
  determined by $\Gamma^{-1}_{\s \si}(\bq, i \Omega_m)=
  -\frac{1}{\lambda_{\s \si}}- \cC_{\s \si}(\bq, i \Omega_m) $, where
  $\cC_{\s \si}(\bq, i \Omega_n)$ is the correlator of the molecular
  (Cooper-pair) field operator $B_{\s \si}(\bq)= \int \frac{d^3
  \bp}{(2 \pi)^3} \; \psipd_\s (\bp+\bq) \psipd_\si (-\bp)$, that to
  leading order in $1/N$ is given by
\begin{widetext}
\begin{align}
\label{zeroth}
\cC_{\s \si}(\bq,i\Omega_m)&= \int_0^\beta d \tau \; e^{i\Omega_m
  \tau} \langle T_\tau \left\{ B^{\vphantom{\dagger}}_{\s \si}(\bq,\tau) B^{\dagger}_{\s \si}(\bq,0)
  \right\} \rangle \notag \\
&=   \frac{1}{\beta} \sum_{\omega_n} \int \frac{d^3
  \bp}{(2 \pi)^3}  \; G^{(0)}_\s(\bp+\bq,i\omega_n+i \Omega_m)
G^{(0)}_\si (-\bp,-i\omega_n)\notag \\
&
=-\int\frac{d^{3}\bp}{\left(2\pi\right)^{3}}
\frac{1-n_F( \xi_{\s, \bp_+})-n_F( \xi_{\si,
    \bp_-})}{i\Omega_m-\xi_{\s, \bp_+}-\xi_{\si,
    \bp_-}},
\end{align}
\end{widetext}
and $\bp_{\pm}=\bp \pm \bq/2$, $\xi_{\s,\bp}=\varepsilon_{\s,
  \bp}-\mu_{\s}$, $n_F(x)={1}/({e^{\beta x}+1})$. An explicit expression for
$\Gamma^{-1}_{\s \si}(\bq,\omega)$ at zero temperature may be found in
the Appendix of Ref.~\onlinecite{nikolic2007}.

For our analysis, we will need the retarded fermionic self-energy
$\Sigma^R_{\s \si}(\bk,\omega)$ at real frequencies and at  finite
temperature. We denote with index $R/A$ the retarded/advanced
functions, {\em i.e.}, functions analytical in the upper/lower
half-planes of the complex frequency. In some cases, it can be
obtained directly from $\Sigma_{\s \si}(\bk,i \omega_{n})$ via a
replacement $i\omega _{n}\rightarrow \omega +i\delta $. However, in
general it is rather difficult to deal with discrete Matsubara
sums. The approach we adopt here is to find the imaginary part of the
retarded self-energy $\TIm \left[\Sigma^R_{\s \si}(\bk,\omega
  )\right]$ and obtain the real part via the Kramers-Kronig relation.

Applying a Lehmann spectral representation
\be
 f(i\omega_{n})=\frac{1}{\pi } \int_{-\infty}^{\infty} dz \frac{ \TIm \left[
f^R(z) \right] }{z-i\omega _{n}},  \label{spectral}
\ee
to (\ref{SE}), we find

\begin{widetext}
\be
 \Sigma_{\s \si}(\bk,i \omega_n )=-\frac{4}{ N \beta}
\int \frac{d^3 \bq}{(2 \pi)^3}
\sum_{ \Omega_m} \int_{-\infty}^{\infty} \frac{dz}{2 \pi}
\int_{-\infty}^{\infty} \frac{dz'}{2\pi} \TIm \left[ \Gamma^R_{\s \si}(\bq,z)
\right] \TIm \left[G^{R (0)}_{\si}(\bq-\bk,z')\right] \times \frac{1}{z-
i \Omega_m} \frac{1}{z' - i \Omega_m +i \omega_n} \label{2.2}
\ee

After summing over the Matsubara frequencies $\Omega_m$
and  performing the standard analytical continuation $i \omega_n \to
\omega + i \delta$, we arrive to
\be
 \Sigma^R_{\s \si}(\bk,\omega)=-\frac{4}{ N} \int \frac{d^3 \bq}{(2 \pi)^3}
\int_{-\infty}^{\infty} \frac{dz}{2 \pi} \int_{-\infty}^{\infty}
\frac{dz'}{2\pi} \frac{n_B( z)+n_F( z')}{ z'-z +\omega
+i \delta} \TIm
\left[ \Gamma^R_{\s \si}(\bq,z) \right]
\TIm \left[G^{R (0)}_{\si}(\bq-\bk,z')\right]  \label{2.2a}
\ee
where $n_B(z)=\frac{1}{e^{\beta z}-1}$. The imaginary part of this
function can be obtained by performing the $z'$ integral to obtain
(denoting $z$ by $\Omega$)
\be
\TIm \left[ \Sigma^R_{\s \si}(\bk,\omega ) \right]=\frac{2}{ N} \int \frac{d^3 \bq}{(2 \pi)^3} \int_{-\infty}^{\infty} \frac{d\Omega}{2\pi}   \TIm
\left[ \Gamma^R_{\s \si}(\bq,\Omega) \right] \TIm \left[G^{R (0)}_{
\si}(\bq-\bk,\Omega-\omega)\right]
\left[n_B(\Omega)+n_F(\Omega-\omega)\right] . \label{2.2b}
\ee
Using the Kramers-Kronig relation
\be
\TRe \left[ f(z) \right]= \frac{1}{\pi} \mathbb{P}
\int_{-\infty}^{\infty} d z' \frac{ \TIm \left[f(z') \right]}{z'-z}
\ee
we obtain the real part of the self energy
\begin{align}
 \TRe \left[\Sigma^R_{\s \si}(\bk,\omega )\right] =\frac{2}{ N} \int
 \frac{d^3 \bq}{(2 \pi)^3} \int_{-\infty}^{\infty} \frac{d\Omega}{2
   \pi} &  \left[  -
\TIm \left[ \Gamma^R_{\s \si}(\bq,\Omega) \right] \TRe
\left[G^{R (0)}_{\si}(\bq-\bk,\Omega-\omega)\right] n_B( \Omega)
\right. \notag  \\
& \left.   + \TRe \left[
\Gamma^R_{\s \si}(\bq,\Omega) \right] \TIm \left[ G^{R (0)}_{\si}
(\bq-\bk,\Omega-\omega)\right] n_F(\Omega-\omega)\right] .
\label{2.2c}
\end{align}

\end{widetext}

As mentioned in the Introduction, we are interested in the
experimentally-relevant case where the number of atoms in each
hyperfine state (species) is independently conserved.  The Fermi
wavevector $\kFS$ for each hyperfine state is related to the
corresponding density by the standard relation $n_\s={\kFS^3}/({6
  \pi^2})$.  Since the atom densities are fixed at $n_\s$, the
nontrivial self-energy corrections modify the dispersions and
therefore lead to shifts in the chemical potentials $\mu_\s$ from
their bare values of $\mu_{\s o}$.

More specifically, the chemical potentials are determined by the
condition $G^{-1}(\kFS,\omega=0)=0$, yielding to lowest order in $1/N$
the relation
\begin{align}
\mu_\s& = \mu_{\s o}+\delta \mu_\s \notag \\
\mu_{\s o}&=\varepsilon_{\sigma, \kFS} \notag \\
\delta \mu_\s & =\TRe \left[\Sigma_\sigma(\kFS,0)\right]. \label{mu}
\end{align}
where $\delta \mu_\s$ are the chemical potential shifts for states
$\s$.  At this point it is important to remark that although
Eq.~(\ref{mu}) is strictly a self-consistent equation, where $\mu_\s$
appears inside the self-energy, it would be overstepping the accuracy
of the $1/N$ expansion to use anything other than the bare values
$\mu_{\s o}$ for the non-interacting Fermi gas at a particular
polarization inside $\Sigma_{\s \si}$.

Another quantity of interest is the two-particle correlation function
\begin{widetext}
\begin{align}
X_{\s \si} (\bq, i\nu_n)
&= -\int_0^\beta  d \tau \int d^3 \br e^{-i \bq\cdot\br
 +i \nu_n \tau} \; \langle T_\tau \left\{
\psid_\s(\br,\tau)\psipd_\si(\br,\tau)\psid_\si(0,0)\psipd_\s(0,0)
\right\}\rangle,
\label{X}
\end{align}
where $\nu_n= 2 \pi n/\beta$ is the bosonic Matsubara frequency. To
order $1/N$, we find for $\s\neq \si$,

\begin{align}
 X_{\s \si} (\bq, i\nu_n)&=
\frac{1}{\beta} \sum_{m} \int \frac{d^3
\bk}{(2\pi)^3}
G_\s(\bk,i\omega_m)
G_\si(\bk+\bq,i\omega_m+i\nu_n)  \times \notag \\
&\left[1+ \frac{1}{N} \frac{1}{\beta} \sum_{\ell} \int \frac{d^3
\bk'}{(2\pi)^3}
G_\s(\bk',i\omega'_\ell)
G_\si(\bk'+\bq,i\omega'_\ell+i\nu_n)
\Gamma_{\si \s}(\bk+\bk'+\bq,i\omega_m+i\omega'_\ell+i\nu_n) \right],
\label{XX}
\end{align}
which, upon using a Lehmann spectral representation and analytic
continuation to real frequencies becomes
\begin{align}
&X_{\s\si}^R(\bq, \nu)=
\int d\omega \int d\omega'
\int \frac{d^3
\bk}{(2\pi)^3}
A_\s(\bk,\omega)
A_\si(\bk+\bq,\omega') \frac{n_F( \omega)-n_F(
  \omega')}{\nu+\omega-\omega'+i \delta}
\notag \\
&+\frac{1}{N} \int \left[\prod_{i=1}^5 d z_i \right] \int \frac{d^3
\bk}{(2\pi)^3} \int \frac{d^3 \bk'}{(2\pi)^3} A_\s(\bk,z_1)
A_\si(\bk+\bq,z_2) A_\s(\bk',z_3) A_\si(\bk'+\bq,z_4) \frac{\TIm
\left[ \Gamma^R_{\si \s}(\bk+\bk'+\bq,z_5) \right]}{\pi}\times
\notag \\
& \frac{1}{z_2-z_1-\nu-i \delta} \frac{1}{z_4-z_3-\nu-i\delta} \left[
\frac{n_{BF}(z_2,z_3,z_5)}{z_5-z_2-z_3} +
\frac{n_{BF}(z_1,z_4,z_5)}{z_5-z_1-z_4} -
\frac{n_{BF}(z_1,z_3,z_5)}{z_5-z_1-z_3-\nu-i\delta} -
\frac{n_{BF}(z_2,z_4,z_5)}{z_5-z_2-z_4+\nu+i \delta} \right],
\label{XXX}
\end{align}
\end{widetext}
where $n_{BF}(x,y,z)=\left[1-n_F(x)-n_F(y)\right] n_B(z)-n_F(x)n_F(y)$, and
\be
A_\s(\bk,\omega)= - \frac{1}{\pi}
\TIm \left[ G^R_{\s\textsc{}} (\bk,\omega) \right]
\label{as}
\ee
is the atomic spectral function.

We will be particularly interested in the case where the state $\ket
3$ is empty, i.e., $\kF_3=0$. In this case, it is useful to
parametrize the two remaining Fermi wavevector $\kF_1$ and $\kF_2$ in
terms of the Fermi energy, $\eF$, and the polarization, $P$. The Fermi
energy is defined as $\eF=\kF^2/(2 m)$, where $n=n_1+n_2 \equiv
\kF^3/(3\pi^2)$. The polarization, a measure of the population
imbalance between states $\ket1$ and $\ket2$, is defined as

\be
P \equiv \frac{n_1-n_2}{n_1+n_2}=
\frac{(\kF_1)^3-(\kF_2)^3}{(\kF_1)^3+(\kF_2)^3}
\ee

The evaluation of Eq.~(\ref{SE}) is meaningful until the condition
$\Gamma^{-1}_{\s \si}(\bq,0)>0$ is violated for some $\bq$, indicating an
instability toward condensation of pairs with this momentum.
At unitarity the instability in $\Gamma_{12}$ occurs at polarization
$P=0.894(8)$, and $q=0.773$. Although this is below the critical value
$P_c=0.93$ that follows from the mean field analysis, it should be
remembered that the transition to the superfluid state is first order
so that the lower value corresponds to a spinodal instability to some
spatially inhomogeneous (FFLO)
state~\cite{sheehy2006,sheehyAOP2007,parish2007,lamacraft2007}.

\section{Fermi liquid properties:  Spectral function}
\label{fermi}

One-particle quantities such as the momentum distributions $n_\s(\bk)$
are conveniently expressed in terms of the fermion spectral functions,
$A_\s(\bk,\omega)$, obtained from the retarded Green's function,
Eq.~\ref{as}.  Note that in the case at hand, the medium and the
interaction are isotropic and therefore the Green's function does not
depend on the orientation of $\bk$. We will therefore drop the vector
notation for $\bk$ from now on.  The momentum distribution,
$n_{\sigma}(k)$ is then given by \be n_{\s}(k) = \int_{-\infty}^\infty
d \omega \; n_F( \omega) A_\s(k,\omega).
\label{nk}
\ee

Near the Fermi surface, we can approximate the Green's function as a
pole plus an incoherent spectrum
\be G^R_{\sigma}(k,\omega)=\frac{Z_{\sigma}}{\omega-E_\sigma(k)+i
\gamma_{\sigma}(k)} + \cdots,
\ee
where $E_\sigma(k)$ is the renormalized spectrum of excitations,
$\gamma_\sigma(k)$ is the quasiparticle decay rate and $Z_\sigma$ is
the residue of the Green's function.  This allows us to characterize
the corresponding strongly-interacting normal state in terms of the
Fermi liquid parameters. The renormalized spectrum is connected to the
real part of the self energy through the relation
\be
\label{EE}
E_{\s}(k)=\varepsilon_{\sigma k} -\mu + \TRe \left[
  \Sigma^R_{\s}(k,E_\s(k)) \right].
\ee
The quasiparticle decay rate is given by $\gamma_\sigma(k)=-\TIm
\left[ \Sigma^R_\sigma \left(k,E_{\sigma}(k) \right)\right]$, whereas
the residue is given by
\be
\label{Z}
Z_\s = \left( 1 - \frac{\partial}{\partial \omega}
  \TRe \left[ \Sigma^R_\sigma(\kFS,\omega)\right]
\right)^{-1}\Biggr|_{\omega=0}.
\ee
The effective mass $m^*_{\sigma}$ of the quasi particle at the Fermi
surface is given by
\be
\label{m*}
\frac{m^*_\s}{m} = \frac{1}{Z_\sigma}
\left( 1 + \frac{m}{k} \frac{\partial}{\partial k} \TRe
\left[ \Sigma^R_\sigma(k,\omega)\right]
    \right)^{-1}\Biggr|_{k=\kFS,\omega=0}.
\ee

We note that Eq.~(\ref{SE}) coincides with the T-matrix approximation,
and therefore will reproduce the known results in the dilute limit
$na^3_{\s \si} \ll 1$ obtained from the pseudopotential
$\frac{4\pi}{m} a_{\s \si}\delta(\br-\br')$.

\subsubsection{Numerical Results}

For $P=0.9$ at unitarity (i.e., $a_{12}=\infty$), the results for the
majority ($\ket1$) and minority ($\ket2$) momentum distribution
functions (Eq. \ref{nk}) are shown in Fig.~\ref{fig:nk}. The Migdal
discontinuity at the Fermi wavevector is a measure of the quasiparticle
residue and correspondingly are given by $Z_1=0.57$ and $Z_2=0.30$. An
interesting feature of this distribution is that $n_\dd(0)\sim 0.45$,
showing strong depletion (from the maximum value of $1$) of the
single-particle states down to zero momentum.

\begin{figure}
\centering \includegraphics[width=0.4\textwidth]{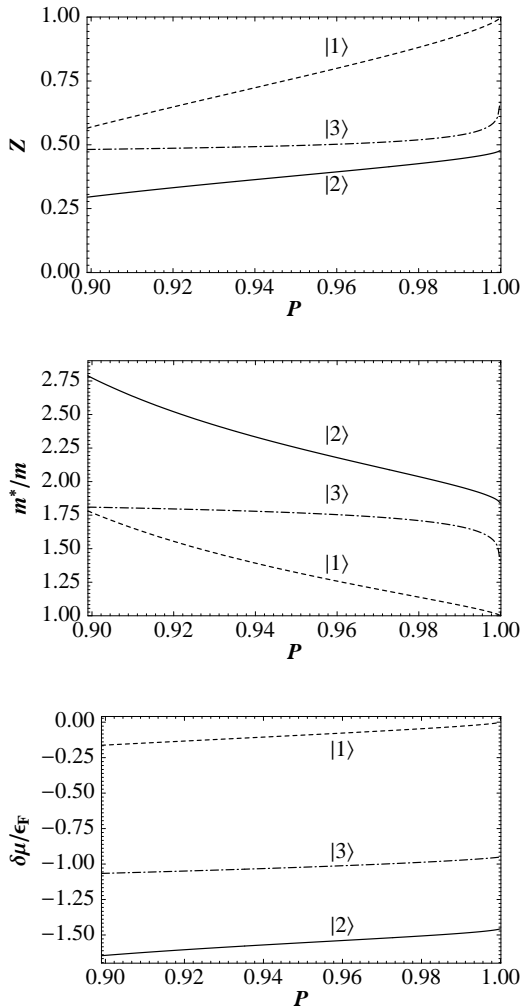}
\caption{Quasiparticle Fermi liquid properties:(A) residue $Z_\s$
  (Eq. \ref{Z}), (B) effective mass $m^*_\s$(Eq. \ref{m*}) and (C)
  the chemical potential shift $\delta\mu_\s $ (Eq. \ref{mu}) of the
  hyperfine state $\ket1$ (dashed line), $\ket2$ (solid line) and
  $\ket3$ (dash-dotted line). }
\label{fig:zm}
\end{figure}

Quasiparticle properties, namely quasiparticle residues, effective
masses and chemical potential shifts, are shown in Fig.~\ref{fig:zm}.
The properties are evaluated at large polarization $P\sim 0.9-1.0$ and
for densities appropriate to $^6$Li at the $B=838$ G resonance.  This
corresponds to $a_{12}=\infty$, $a_{13}=-3288 a_o$ and $a_{23}=-16080
a_o$ where $a_o$ is the Bohr radius ($a_o=0.0529177$nm). In order to
make contact with the experiment of Schunck {\it et
  al.}~\cite{schunck2007}, we set the atomic density such that $\kF
a_{13} =-3.3$ (which directly implies $\kF a_{23}=-16.1$).  However,
we stress that, given that the hyperfine state $\ket3$ is empty, the
single particles properties of the hyperfine states $\ket1$ and
$\ket2$ do not depend on the values of $a_{23}$ and $a_{13}$. In this
sense their properties are universal and our numerical result are
applicable to other fermionic resonant atoms, e.g., $^{40}$K. On the
other hand, the properties of the hyperfine state $\ket3$ are not
universal in the sense that they depend on the interaction couplings.

Using the formula Eq.~(\ref{Z}), we evaluate the quasiparticle
residues $Z_\s$ at zero temperature. At full polarization ($P=1$), the
residue $Z_1=1$ is maximum since the state $\ket2$ is empty. On the
other hand, the residue $Z_2=0.47$ is much smaller than unity,
indicating a large renormalization of atoms in state $\ket2$ due their
interaction with atoms in state $\ket1$.

The binding energy of the minority carrier is $\mu_{2}= -1.46 \eF$,
which can be compared to the Monte Carlo calculation of
Ref.~\onlinecite{lobo2006} that found $\mu_2=-0.93 \eF$.

It is important to note here that previous computations \cite{lobo2006} of the
dispersion spectrum of the minority spin have all been done for
the case of a single minority spin, in which case the problem is analogous
to a Kondo/X-ray edge problem. Further, the dispersion spectrum is
measured directly at ${\bf k} = 0$. In contrast, we are considering here 
the case of a finite density of minority spins, and the Fermi liquid properties
near $k=k_{F2}$, with $|k-k_{F2}| \ll k_{F2}$. Even as $k_{F2} \rightarrow 0$, this regime
is distinct from the single minority spin case. With unitary interactions, it is not
clear how the two regimes will connect. We leave this important issue open
for future work.

\section{Radio Frequency  Spectroscopy}
\label{radio}

We turn next to the RF spectroscopy of
Refs.~\onlinecite{chin2004,schunck2007}.  The RF spectroscopy is a
technique used to probe atomic correlation by exciting atoms from
occupied hyperfine states to another  (usually empty) reference
hyperfine state. As we will see below, the RF probe provides
valuable information about single particle excitations.

In the RF experiments of Schunck \emph{et al.}, one focuses on three
different atomic hyperfine states of the $^6$Li atom.  The two lowest
states, $\ket1$ (majority) and $\ket2$ (minority), are populated and
responsible for the superfluid correlations. The higher state,
$\ket3$, is empty initially, and is used as a probe of atomic
correlations in states $\ket1$ and $\ket2$.  An RF field, at
sufficiently large frequency, is used to drive atoms from state
$\ket2$ to state $\ket3$.

In general the signal is expressed as a two-particle quantity.  We
calculate the rate for this process, described by a tunneling
Hamiltonian that couples one of the species to a reference state
$\ket3$ with a frequency $\omega$ detuned from the bare atomic
transition.

The RF field induces a transition between atomic levels primarily
through the electronic spin. In the AC field of interest, the
rotating wave approximation is used to described the tunneling
Hamiltonian in terms of transfer matrix elements $\TU_{\bk,\bp}$
between $\ket2$ to $\ket3$ state, i.e.,
\be
\ham_T  =\int \frac{d^3 \bk}{(2 \pi)^3}\frac{d^3 \bp}{(2 \pi)^3}
\left[\TU_{\bk \bp}\,e^{-i \omega_L t}\psid_{3}( \bk) \psipd_{2}(\bp)
  +h.c. \right]
\ee
For plane wave states, the tunneling matrix elements are
$\TU_{\bk \bp} = \bar {\TU
} (2 \pi)^3 \delta({\bq}_L+{\bk}-{\bp})
.$
Here $q_L \approx 0$ and $\omega_L$ are the momentum and energy of the
RF field. The conservation of momentum in tunneling is to be
contrasted with the analogous Hamiltonian used to model tunneling in
superconducting-metal junction.

The RF current is defined as $\hat{I}_{3}
=\dot{N}_3=i[\ham_T,\hat{N}_3]$ where $\hat{N}_3=\int d^3 \br \;
\psid_3(\br) \psipd_3(\br)$ is the number operator of the hyperfine
state $\ket3$. Thus the current operator is
\be
\hat{I}_3(t)=-i
\bar{\TU} \int \frac{d^3 \bk}{(2 \pi)^3} \left[ e^{-i \omega_L t}
  \psid_3(\bk) \psipd_2(\bk) - h.c. \right]
\ee
where as an accurate approximate we have neglected the extremely low
RF photon momentum and set $\bq_L=0$. A standard linear response
analysis gives
\be I_3(\nu) \equiv \langle \hat{I}_3(\nu) \rangle=2
\bar{\TU}^2\TIm \left[ X^R_{32} ({\bf 0},\nu) \right],
\label{RF}
\ee
where $\nu = \omega_L+\omeps_2-\omeps_3-\mu_2+\mu_3$ is the effective
detuning. The retarded response function $X^R_{32}({\bf 0},\nu)$ is
obtained from the analytic continuation of Eq.~\ref{X}, giving
\begin{align}
\TIm \left[ X^R_{32}({\bf 0},\nu)\right] = \TIm \left[ X^{(a)
R}_{32}({\bf 0},\nu) \right] + \TIm \left[ X^{(b) R}_{32}({\bf
0},\nu) \right],
\end{align}
where
\begin{widetext}
\begin{align}
\TIm&\left[ X^{(a) R}_{\s \si}({\bf 0},\nu) \right]= -\pi \int
\frac{d^3 \bk}{(2\pi)^3} \int_{-\infty}^\infty d \omega \left[
n_F(\omega) -n_F(\omega+ \nu) \right] A_\s(\bk,
\omega) A_\si(\bk,\omega+\nu)\label{xa}
\\
\TIm &\left[ X^{(b) R}_{\s \si}({\bf 0},\nu) \right]=
-\frac{\pi}{N} \int \frac{d^3 \bk}{(2\pi)^3} \int \frac{d^3 \bk'}{(2\pi)^3} \int_{-\infty}^\infty
 d \omega \int_{-\infty}^\infty d \omega'\; \times \notag  \\
\Big\{&-n_{BF}( \omega+ \nu, \omega'+ \nu,
\omega+\omega'+\nu) G_\s(\bk, \omega) A_\si(\bk,\omega+\nu)
G_\s(\bk', \omega') A_\si(\bk',\omega'+\nu) \Pi_{\si \s} (\bk+\bk', \omega+\omega'+\nu)  \notag \\
&+n_{BF}( \omega, \omega',\omega+\omega'+\nu) A_\s(\bk, \omega) G_\si(\bk,\omega+\nu)
A_\s(\bk', \omega') G_\si(\bk',\omega'+\nu) \Pi_{\si \s} (\bk+\bk', \omega+\omega'+\nu) \notag \\
&+2 \left[n_F(\omega) -n_F(\omega+ \nu)\right]A_\s(\bk,\omega)A_\si(\bk,\omega+\nu) \times  \notag \\
&\,\,\,\,\,\,\,\,\,\,\,
\big[n_B( \omega'+\omega+\nu) G_\s(\bk',\omega') G_\si(\bk',\omega'+\nu) \Pi_{\si \s}(\bk+\bk',\omega+\omega'+\nu) \notag \\
&\,\,\,\,\,\,\,\,\,\,\,\,\,-n_F(\omega') A_\s(\bk',\omega') G_\si(\bk',\omega'+\nu) 
\Gamma_{\si \s}(\bk+\bk',\omega+\omega'+\nu) \notag \\
&\,\,\,\,\,\,\,\,\,\,\,\,\,-n_F(\omega'+\nu) G_\s(\bk',\omega') A_\si(\bk',\omega'+\nu)
 \Gamma_{\si \s}(\bk+\bk',\omega+\omega'+\nu) \big] \Big\}
\label{xb}
\end{align}
where $\Pi_{\s \si}(\bq,\nu)=
-\pi^{-1}\TIm \left[ \Gamma_{\s \si}(\bq,\nu) \right]$.

An explicit expression for $\Gamma^{-1}_{\s \si}(\bq,\nu)$ at zero
temperature may be found in the Appendix of
Ref.~\onlinecite{nikolic2007}. We note that in the limit of a vanishing interaction between states $\ket2$
and $\ket3$, i.e., for $a_{23}=0$, there are no vertex corrections to
the $X^R_{32}({\bf 0},\nu)$ correlator, and we find
\begin{align}
\TIm \left[ X^R_{32}({\bf 0},\nu)\right]& = \TIm \left[ X^{(a)R}_{32}({\bf 0},\nu) \right] \notag \\
   &=
-\pi \int \frac{d^3 \bk}{(2\pi)^3} \int_{-\infty}^\infty d \omega
\left[ n_F(\omega) -n_F(\omega+ \nu) \right] A_3(\bk, \omega) A_2(\bk,\omega+\nu),
\end{align}
with all the interactions and corresponding $1/N$ corrections entering
only through the atomic spectral functions $A_{2,3}(\bk,\omega)$.

\end{widetext}

\subsubsection{Numerical Results}

We evaluate the RF spectrum for $^6$Li at the resonance from the above
expressions, displaying the result in Fig.~\ref{fig5}. As we
mentioned in the previous section, interactions between particles
introduce self-energy corrections to the fermions.
 In the calculation of the RF spectrum, it is important to
consider all scattering lengths because the spectral weight of the
states $\ket2 $ and $\ket3 $ are affected by all interactions.
Following the formula Eq.~(\ref{xa}), we can evaluate the
contribution $X^{(a)}_{32}({\bf 0},\nu)$. We find that the bosonic
propagators $\Gamma_{13}({\bf q},w)$ and $\Gamma_{23}({\bf q},w)$
have sharp poles at low momenta and positive energies. These
positive energy ``Aleiner-Altshuler'' poles are analogous to
the bifermion mode discussed in the context of weak superconductors
above the paramegnetic limit \cite{Aleiner1997}. The sharp poles
provide an additional spectral weight of the states $\ket 2$,
$\ket 3$ but are not found to be significant contributions given
the large energy gap associated with these modes.

The interaction between states $\ket 2$ and $\ket 3$ introduces a
vertex correction term, $X^{(b)}_{32}({\bf 0},\nu)$, to the RF
spectrum. Using the formula Eq.~(\ref{xb}) and the condition that
the reference state $\ket 3$ is empty, we find that the vertex
correction is not significant in the limit $P \sim 1$. For
example, for polarization $P = 0.97$ the vertex correction
is less than 0.1 $\%$, as shown in Fig.~\ref{vertex097}.

\begin{figure}
\centering
\includegraphics[width=0.45\textwidth]{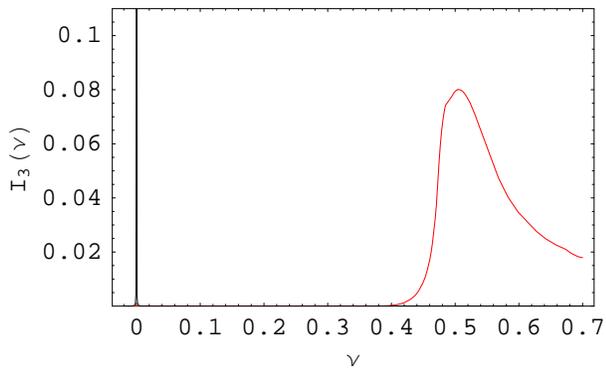}
\caption{RF spectrum for polarization $P=0.93$:
 the intensity
$I_3(\nu)$ (arbitrary units)
  vs. the detuning from the resonance frequency $\nu$ measured
  in units of the Fermi energy $\eF$. The peak frequency relative to
non-interacting line (black line) is smaller than for the $P=0.97$ case shown in
Fig.~\ref{fig5}. The
linewidth broadens for lower polarizations due to the increase in pairing fluctuations.} \label{figRF}
\end{figure}

\begin{figure}
\centering
\includegraphics[width=0.45\textwidth]{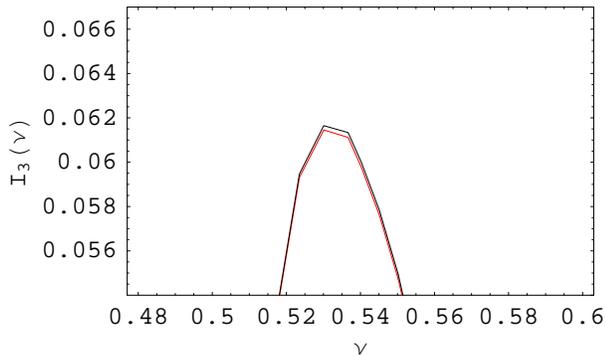}
\caption{Vertex correction to the RF spectrum for $P=0.97$. The
axes are the same as in Fig.~\ref{figRF}. The red line includes the vertex correction,
while the black line does not. The contribution of 
the correction $X^{(b)}_{32}({\bf 0},\nu)$ is less than 0.1 $\%$. } \label{vertex097}
\end{figure}

In Figs.~\ref{figRF} and \ref{fig5}, we show the RF spectrum for
the imbalanced polarization $ P=0.93$  and $P=0.97$, respectively.
Both RF signal peaks are positioned near  $\Delta \sim 0.55~\eF$
which is nearly equal to the chemical potential difference between
level $\ket 2$ and $\ket 3$ (see Fig. \ref{fig:zm}). The results
for the peak position compare favorably to the recent MIT
experiment of Schunck {\it et al.}~\cite{schunck2007} where they
found $\Delta\sim 0.38~\eF$ at finite temperature $T/T_{F}\sim
0.08$. The position of the peak for $P=0.93$ is closer to the
non-interacting peak than $P=0.97$ case, and the increased pairing
fluctuations at $P=0.93$ further broaden the linewidth.

\section{Conclusion}
\label{conclusion}
We have studied the normal state of a resonantly interacting, three
component Fermi gas at large population imbalance. To this end, we
utilized a well-controlled large $N$ (atom flavor) expansion, with
$1/N$ as the small expansion parameter and showed that it provides a
systematic way to treat the strong interactions characteristic of the
unitary scattering point of a Feshbach resonant system.  
Although the accuracy of the $1/N$ expansion in the limit
$N\to 1$ remains unproven in general, it has provided reasonable 
estimates in other settings.
Our main aim was to provide a qualitative interpretation of 
recent RF experiments~\cite{schunck2007}, and to emphasize the importance
of measuring quasiparticle properties to determine the nature of the
ground state of imbalanced Fermi gases.
 Performing the analysis to  leading order in $1/N$, we computed the atomic
spectral functions and the momentum distribution functions with the
characteristic Migdal discontinuity, thereby showing that the normal
state is a conventional Fermi-liquid, albeit strongly renormalized. We
then applied this formalism to analyze the RF excitation spectrum
studied experimentally in Ref.~\onlinecite{schunck2007}. Our
conclusion is that, indeed, the observed phenomenology can be well
understood in terms of the strongly interacting, but conventional
Fermi-liquid picture without resorting to exotic interpretations, such
as, for example, pairing without condensation.

\paragraph*{Note added:} Recent experiments by the MIT group \cite{ketterle2008}
have been interpreted in a physical picture consistent with that
presented above, in contrast to their earlier interpretation of 
previous experiments \cite{schunck2007}.

\acknowledgements
We thank W.~Ketterle and Y. Shin for useful discussions.
This research was supported in part by the National Science Foundation
under Grants No. DMR-0321848 (MV, DS, LR), DMR-0757145 (EGM,SS),
and at the Kavli Institute for Theoretical Physics under grant PHY05-51164.
EGM is also supported in part by the Samsung Scholarship.

\end{document}